# The dimensionality of a coiled helical coil

Subhash Kak
Oklahoma State University, Stillwater, OK 74078, USA
Email: subhash.kak@okstate.edu
ORCID: 0000-0001-5426-9759

**Abstract**
The helical coil is ubiquitous in biological and natural systems and often it is the basic form that leads to complex structures. This paper considers the question of its dimensionality in biological information as the helical coil goes through recursive coiling as happens to DNA and RNA molecules in chromatin. It has been shown that the dimensionality of coiled coils is virtually equal to $e$. Of the three forms of DNA, the dimensionality of the B-form is nearest to the optimal value and this might be the reason why it is most common.

*Keywords*: helical coil, fractals, artificial helical structures, noninteger dimensionality

**Introduction**

Helical or coiled structures are seen in proteins and nucleic acids, and at the macroscopic level in the case of the spiral anatomy of the heart muscle bands or the helical twisting of the umbilical cord [1]. Artificial helical structures ranging from nano- to macro-scales have been developed as in nanosprings made of zinc oxide, and helical microtubules of graphitic carbon [2]. In cosmology, we see them in the agglomeration of galactic nebulae and in plasma jets. In general, the dimensionality of scale invariant systems turns out to have noninteger or fractional value [3].

In the case of chromatin, we see that the coiled helical shapes of DNA and RNA give rise to a structure that has noninteger dimensionality [4], and thus the question of the characterization of different kinds of helical coils becomes important. A recent study considered the dimensionality of the related circular and spiral phenomena [5], which was motivated by the search for structures that are optimal in an information theoretic sense [6][7][8].

It should be noted that the use of optimality arguments is a powerful method to obtain understanding of biological phenomena [9][10][11], and it has been used to explain several hitherto unexplained features of the genetic code [12][13][14] and of neural structures [15][16]. Specifically, this approach has been shown to explain the irregular mapping of the codons into amino acids, and one may hope that this approach will be useful in elucidating other aspects of the genetic code as well [17][18].

In considering a helical coil, its structure is apparent only if it is seen from near, because from a distance it looks like a plain string with dimensionality of 1. When we come close, its loops come into view, and it appears three-dimensional. So, its *apparent dimensionality* (D) is dependent on the information available to the observer.

The D value will make sense for structures defined in a recursive sense. In the continuous case, one may wish to associate it with how the coil is further looped at next levels. A well-known examples of recursive coiling is the structure of wool which is proteins together with lipids with a unique surface structure of overlapping cuticle cells (scales) that anchor the fiber into the sheep's skin.

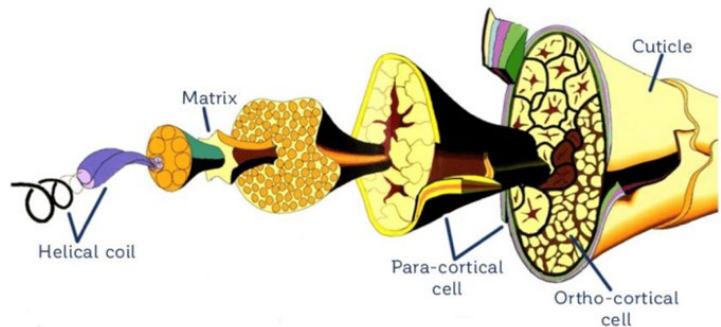

Figure 1. Multilayered coiled structure of wool

The cells have a waxy coating, making the wool water repellent whilst tiny pores that allow water vapor to pass through the wool fiber, and the scales overlap like tiles on a roof. The cortex is the middle layer of the wool fiber that is made up of long, twisted chains of keratin, and it contains many small, air-filled spaces that help to trap heat and insulate the body. The helical coil, which is surrounded by the matrix, is tightly coiled structure and the coils help to distribute stress and tension across the fiber.

Each chromosome is a physical structure formed by supercoiling of the DNA round the scaffold proteins. The DNA coils, then folds back on itself, then coils again until each DNA molecule is so tightly coiled up that a visible chromosome appears in the nucleus.

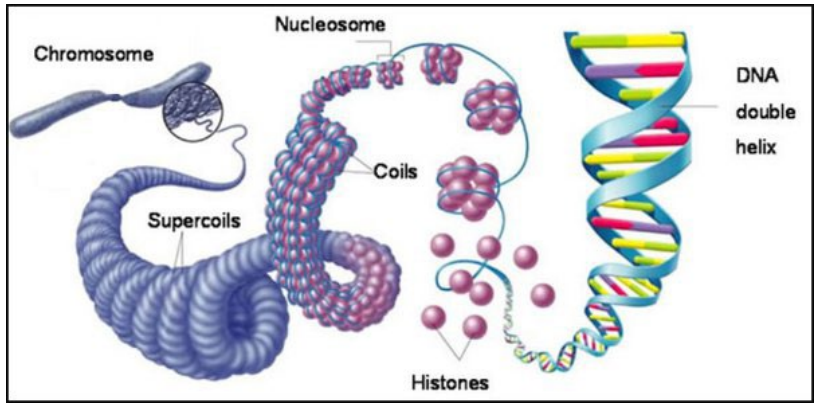

Figure 2. The DNA coils, then folds back on itself, then coils again (PMG Biology)

In this paper, which is a successor to the earlier study on the dimensionality of genetic information [13], we consider the question of the dimensionality of helical coils in its most basic form and show that it is optimal. Certain variants of this basic form that are not optimal are also considered.

**The length of the continuous curve**

Consider a self-similar system which is scale invariant up to a certain minimum size. Let the system size be 1.

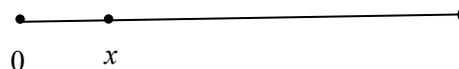

Figure 3. A one-dimensional system of unit size

The system may be taken to have self-similar features at sizes 1, 1/2, 1/4, 1/8, … and so on until the minimum scale associated with the system.

Let us first consider a system of sinusoids in nested levels, which can model general classes of patterns. Since we can represent features as a sum of sinusoids, we shall for simplicity consider the simple case of a single frequency (to which other frequencies may be added to represent any specific shape).

Consider a frequency of $2\pi$ radians. The function $f_1$ at the point $x$ will be

$$f_1(x) = g \sin 2\pi x \qquad (1)$$

Likewise, the functions associated with smaller scales will be, recursively:

$$f_2(x) = \frac{g}{2} \sin 4\pi x \qquad (2)$$

$$f_3(x) = \frac{g}{4} \sin 8\pi x \qquad (3)$$

and so on.

The cumulative function will thus be

$$F(x) = \sum_i f_i(x) = \sum_{i=1}^{N} \frac{g}{2^{i-1}} \sin 2^i \pi x \qquad (4)$$

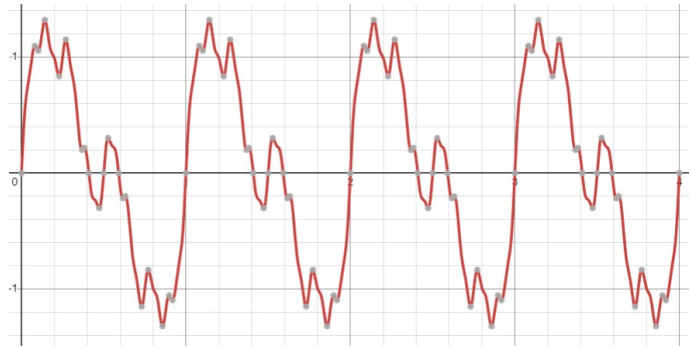

Figure 4. The self-similar sinusoidal function

If, instead of (1), we consider the function $b(m + 2^{(n-1)}) = 2^n + 1/2[b(m) + b(m + 2^n)]$, then we will get the Blancmange Curve instead of (4). The $d$th iteration contains $2^d + 1$ points.

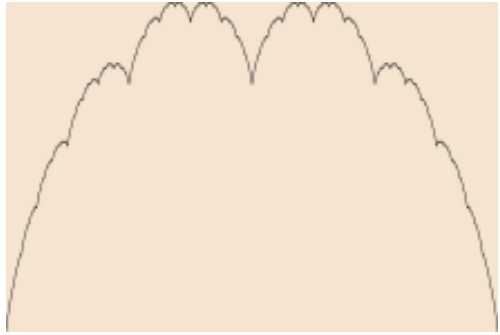
Figure 5. The Blancmange curve

The apparent dimensionality of the curve of Figure 1 may be defined in terms of the excess ratio, which seems intuitive so long as the excess is small.

The arc length ratio for the sine function from 0 to $\frac{\pi}{2}$ equals

$$\frac{2}{\pi}\int_0^{\pi/2} \sqrt{1+g^2\cos^2\theta}\ d\theta \qquad (5)$$

As one can check, when $g=0$, the ratio is 1. We can write $1+g^2\cos^2\theta$ as $1+g^2-g^2\sin^2\theta$ so that equation (5) becomes

$$\frac{2}{\pi}\int_0^{\pi/2}\sqrt{1+g^2-g^2\sin^2\theta}\ d\theta$$

$$=\frac{2}{\pi}\sqrt{1+g^2}\int_0^{\pi/2}\sqrt{1-\frac{g^2}{1+g^2}\sin^2\theta}\ d\theta$$

Note that

$$E(k)=\int_0^{\pi/2}\sqrt{1-k^2\sin^2\theta}\ d\theta,\ -1<k<1$$

is the complete elliptic integral of the second kind, which is mapped below for easy reference:

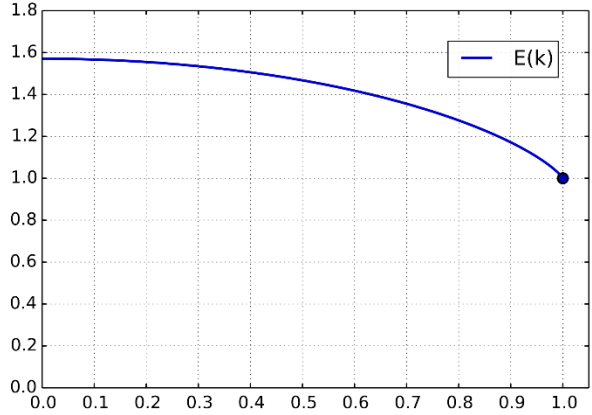

Figure 6. Sketch of the complete elliptic integral of the second kind

We can write the length formula as:

$$L = \frac{2}{\pi}\sqrt{g^2 + 1}\, E\left(\frac{g^2}{g^2+1}\right) \qquad (6)$$

Thus if $g = 0.1$, then L is $\frac{2}{\pi}\sqrt{1.01}E(\frac{0.01}{1.01}) \approx 1.0039$. Some selected values are given in the table below. The dimensionality, D, is obtained by the equation $\frac{\ln 2L}{\ln 2}$ since our process considers recursive doubling.

Table 1. Fractal dimensionality for some values of g for simple sine

| g | L | D=$\frac{\ln 2L}{\ln 2}$ |
|---|---|---|
| 0 | 1 | 1.0 |
| 0.1L | 1.0039 | 1.0056 |
| 0.2L | 1.0119 | 1.0170 |
| 0.3 L | 1.0411 | 1.0580 |
| 0.4L | 1.0708 | 1.0976 |
| L | 1.3207 | 1.4005 |

**The helical coil**

Consider the helical coil shown in Figure 5.

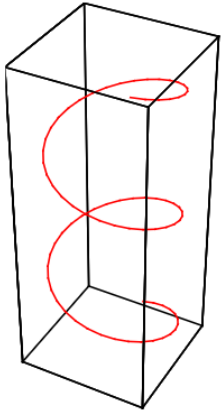

Figure 7. A helical coil

Let the radius of the helix be $r$ and so the location of the point on the curve will be in terms of coordinates

$$x = r \cos t$$
$$y = r \sin t$$
$$z = ct$$

where $t \in (0, 2\pi)$, and so $2\pi c$ gives the vertical separation of the helix's loops. The arc length is given by

$$s = \sqrt{r^2 + c^2}\, t \qquad (7)$$

**A discrete helical coil**

Three major forms of DNA are double stranded and connected by interactions between complementary base pairs. These are A-form, B-form, and Z-form DNA. The dimensions of B-form (the most common) of DNA are 0.34 nm between base pair, which is 3.4 nm per turn with about 10 base pairs per turn, and about 2.0 nm in diameter. The B-form is the classic, right-handed double helical structure.

A thicker right-handed duplex with a shorter distance between the base pairs has been described for RNA-DNA duplexes and RNA-RNA duplexes. This is called A-form nucleic acid. This has 11 base pairs per turn with a length of 2.86 nm. The helix width of A-DNA is 2.3 nm.

A third form of duplex DNA has a strikingly different, left-handed helical structure, with 12 base pairs per turn with a length of 4.56 nm. The distance between two adjacent bases is 0.37 nm and the helix width is 1.8 nm.

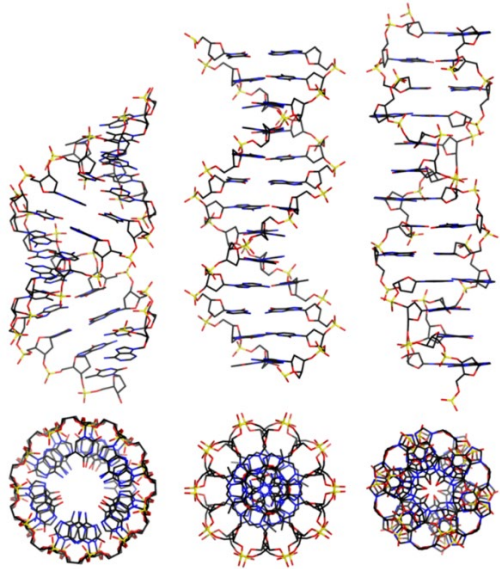

Figure 8. B-form (left), A-form (middle), and Z-form DNA (CC BY-SA 4.0)

The length in one turn for the B-form DNA will be from the vertical length of 3.4 nm and the diameter of 2 nm that equals a circumference of $2\pi$ nms.

$$L_B = 2\pi\sqrt{1 + 3.4^2} \approx 22.26$$

The straight line length of 3.4 nm is stretched to 22.26, or it is stretched by a factor of 22.26/3.4 = 6.547. Assuming this to happen recursively, the dimensionality, D, of the recursively defined B-form must be taken to be:

$$D_B(x) = \frac{\ln 6.547}{\ln 2} \approx 2.7108$$

It is significant that this value s virtually identical to the optimal dimensionality of $e$.

The corresponding calculation for A-form DNA is

$$L_A = 2\pi\sqrt{(1.15^2 + (2.86)^2)} \approx 19.37$$

Normalizing it with respect to the base of 2.86, we get 6.77. Thus

$$D_A(x) = \frac{\ln 6.77}{\ln 2} \approx 2.759$$

The calculation for Z-form DNA is likewise

$$L_Z = 2\pi\sqrt{(0.9)^2 + (4.56)^2} \approx 29.20$$

Normalizing it with respect to the base of 4.56, we get 6.40. Thus

$$D_A(x) = \frac{\ln 6.77}{\ln 2} \approx 2.678$$

It is significant that the dimensionality of the commonly occurring B-form is nearest to the optimal value of *e*.

**Conclusions**

The helical coil it is the basis form that leads to complex structures in many natural and biological systems. This paper computed the dimensionality of the coiled coil for B-form DNA and showed that it is closest to the optimal value of *e* as compared to A- and Z-form DNA. Such an analysis can be extended to artificial helical structures that have been proposed for nanomaterials.